%% file: b2n_xlnu.tex
\documentclass[preprint,prd,tightenlines,superscriptaddress]{revtex4-2}

\usepackage{graphicx} 
\usepackage{dcolumn}  
\usepackage{colordvi}
\usepackage{color}
\usepackage{epstopdf}
\usepackage{amssymb}
\usepackage{amsmath}
\usepackage{url}
\graphicspath{{ps}}
\usepackage{hyperref}
\usepackage{tabularx}
\usepackage{gensymb}
\usepackage{cancel}
\usepackage{enumitem}
\usepackage[separate-uncertainty = true]{siunitx}

\makeatletter
\def\p@subsection{}
\makeatother

\input ./belle2sym.tex

\begin{document}

\def\belletwo {\it {Belle II}}
\def\btoxclnu {$B\to X_c\ell\nu_\ell$}
\def\btoxcenu {$B\to X_c e\nu_e$}
\def\btoxcmunu {$B\to X_c\mu\nu_\mu$}
\def\samplesize{\SI{62.8}{\per\femto\barn}}
\def\samplesizeOffRes{\SI{9.2}{\per\femto\barn}}
\def\MCsize{\SI{200}{\per\femto\barn}}
\newcommand{\NBB}{\ensuremath{N_{B \bar B}=  \left(68.21 \pm 0.06(\text{stat})\pm 0.75(\text{sys})\right)\times 10^6}}

\vspace*{-3\baselineskip}
\resizebox{!}{3cm}{\includegraphics{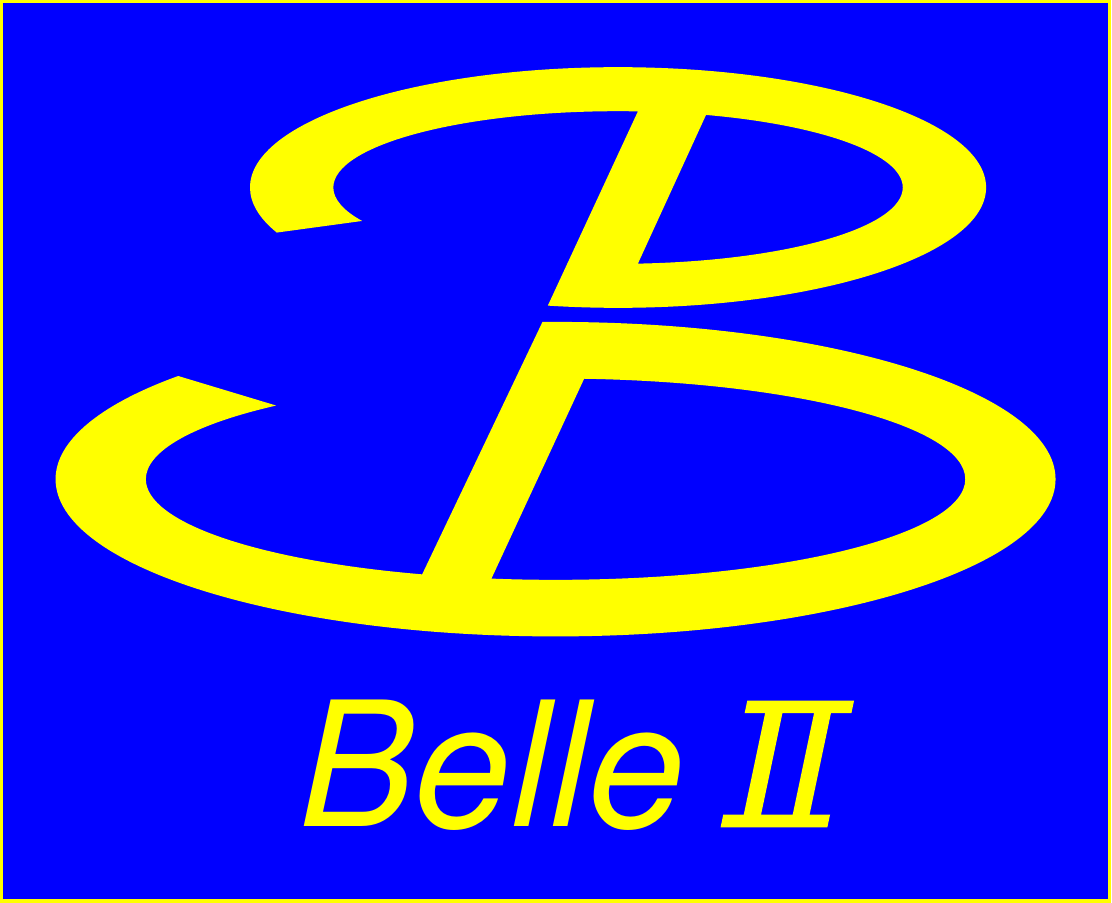}}

\vspace*{-5\baselineskip}
\begin{flushright}
BELLE2-CONF-PH-2021-012\\
\today
\end{flushright}

\title { \quad\\[0.5cm] Measurement of the inclusive semileptonic $B$~meson branching fraction in \samplesize\ of Belle II data }

\input{authors-conf2021}

\begin{abstract}
We report a measurement of the branching fraction of inclusive semileptonic $B$~meson decays \btoxclnu\ in  $\Upsilon(4S)\to B\bar B$~data recorded by the Belle II experiment at the SuperKEKB asymmetric-energy $e^+e^-$ collider and corresponding to \samplesize\ of integrated luminosity. Only a charged lepton (electon or muon) is reconstructed and the signal yield is determined from a fit to the lepton momentum distribution in the center-of-mass frame of the colliding beams. Averaging the result in the electron and muon channels, we find ${\cal B}(B\to X_c\ell\nu_\ell) = (9.75\pm 0.03(\text{stat})\pm 0.47(\text{sys}))\%$.
\end{abstract}


\maketitle

{\renewcommand{\thefootnote}{\fnsymbol{footnote}}}
\setcounter{footnote}{0}

\newpage

\section{Introduction}

The magnitude of the Cabibbo-Kobayashi-Maskawa (CKM)~\cite{Cabibbo:1963yz,Kobayashi:1973fv} matrix element $|V_{cb}|$ squared determines the transition rate of $b$ into $c$ quarks. The precise knowledge of this fundamental parameter of the Standard Model (SM)~\cite{Pich:2007vu} is crucial for the ongoing precision $B$~physics programme at the Belle II experiment and elsewhere. The CKM element $|V_{cb}|$ is measured from semileptonic $B$~meson decays $B\to X_c\ell\nu_\ell$, where $X_c$ is a hadronic system with charm, $\ell$ is a light charged lepton (electron or muon) and $\nu$ is the associated neutrino. These determinations can be $\emph{inclusive}$, {\it i.e.}, sensitive to all $X_c\ell\nu_\ell$ final states within a given region of phase space, or $\emph{exclusive}$, {\it i.e.}, based only on a single $b\to c$ semileptonic mode such as $B\to D^*\ell\nu$ or $B\to D\ell\nu$. Pursuing both approaches is important as the two avenues involve different theoretical and experimental uncertainties and consistency between both is a powerful consistency check of our understanding. However, inclusive and exclusive measurements of $|V_{cb}|$ have been at odds for many years now, an issue which is often referred to as the \emph{inclusive vs.\ exclusive problem}~\cite{Amhis:2019ckw}.

In this paper we describe a measurement of the inclusive semileptonic branching ratio based on the Belle II data collected in the years 2019 and 2020 equivalent to \samplesize. The paper is organized as follows: Sect.~\ref{sec:2} describes the collision data and simulated data samples used in this analysis. Sect.~\ref{sec:3} introduces our experimental procedure. Finally, Sect.~\ref{sec:4} contains all results and the analysis of systematic uncertainties.

\section{The Belle II detector and data sample} \label{sec:2}

The Belle~II detector~\cite{Abe:2010sj} operates at the SuperKEKB asymmetric-energy  electron-positron collider~\cite{Ohnishi:2013fma}, located at the KEK laboratory in Tsukuba, Japan. The detector consists of several nested detector subsystems arranged around the beam pipe in a cylindrical geometry. The innermost subsystem is the vertex detector, which includes two layers of silicon pixel detectors and four outer layers of silicon strip detectors. Currently, the second pixel layer is installed in only a small part of the solid angle, while the remaining vertex detector layers are fully installed. Most of the tracking volume consists of a helium and ethane-based small-cell drift chamber (CDC). Outside the drift chamber, a Cherenkov-light imaging and time-of-propagation detector provides charged-particle identification in the barrel region. In the forward endcap, this function is provided by a proximity-focusing, ring-imaging Cherenkov detector with an aerogel radiator. Further out is the ECL electromagnetic calorimeter, consisting of a barrel and two endcap sections made of CsI(Tl) crystals. A uniform 1.5~T magnetic field is provided by a superconducting solenoid situated outside the calorimeter. Multiple layers of scintillators and resistive plate chambers, located between the magnetic flux-return iron plates, constitute the $K_L$ and muon identification system (KLM).

The data used in this analysis were collected between March 2019 and July 2020 and correspond to \samplesize\ of integrated luminosity on the $\Upsilon(4S)$~resonance (10.58~GeV) and \samplesizeOffRes\ of integrated luminosity below the $\Upsilon(4S)$~resonance (10.52~GeV), referred to as {\it off-resonance data}. Collected data sample contains \NBB\ $\Upsilon(4S)\to B\bar B$ events as determined from a fit to event-shape variables~\cite{BELLE2-NOTE-PL-2020-006}. In addition, we use Monte Carlo (MC) simulated events equivalent to \MCsize\ throughout this analysis. These include a sample of $\Upsilon(4S)\to B\bar B$~events in which $B$~mesons decay generically, generated with EvtGen~\cite{Lange:2001uf} and a sample of continuum $e^+e^-\to q\bar q$~events ($q = u, d, s, c$) simulated with KKMC~\cite{ref:kkmc}, interfaced with PYTHIA~\cite{Sjostrand:2007gs}. The $B\bar B$~sample includes semileptonic $B$~meson decays $B\to X\ell\nu$, where $X$ can be a hadronic system with and without charm. The latter is modeled by a mixture of exclusive modes ($X_u$ can either be charged or neutral $\pi$, $\rho$, $\omega$ or $\eta$) and an inclusive model~\cite{fazio}. Full detector simulation based on GEANT4~\cite{ref:geant4} is applied to MC~events. The lepton reconstruction efficiencies and the hadron misidentification rates in simulation are adjusted to match the real performance of the Belle II lepton identification system.

Both data and simulated events are analysed with Belle II analysis software framework (BASF2)~\cite{Kuhr:2018lps}. Hadronic events are selected and backgrounds coming from quantum electrodynamic processes (low multiplicity events) are reduced by requiring more than three charged tracks in a single event, the total energy of the reconstructed charged and neutral particles above 4~GeV and a ratio $R_2$ of the second to the zeroth Fox-Wolfram moment below 0.4~\cite{Fox:1978vu}.

\section{Experimental Procedure} \label{sec:3}

\subsection{Reconstruction}

We require charged particle tracks to originate from the interaction point (IP): The distance of closest approach between each track and the interaction point is required to be less than 2~cm along the $z$ direction (parallel to the beams) and less than 0.5~cm in the transverse $r-\phi$ plane. We further require charged particles to be within acceptance of the central drift chamber (CDC) and to have transverse momentum above 100~\mevc.

In the next step, we identify charged lepton candidates (electrons or muons). The particle's center-of-mass (c.m.) momentum $p_\ell^*$ must lie in the range between 0.4 and 2.5~\gevc. Electrons are identified based on their energy and shower shape in the ECL calorimeter. Muons are identified using information from the instrumented return yoke KLM. We require the lepton candidates to have momenta in the laboratory frame within the range of $p_\ell \in [0.4,2.5]$~\gevc and polar angle $\theta_e \in [0.22, 2.71]$~rad for electrons and $\theta_{\mu} \in [0.4, 2.6]$~rad for muons. We veto charged leptons from $J/\psi$~decays or from photon conversion. Each lepton candidate is combined with an oppositely charged particle and two regions of invariant mass $M(\ell^+\ell^-)$ are excluded -- the interval $[3.0,3.14]$~\gevcc for electrons and $[3.04,3.14]$~\gevcc for muons. Photon conversions to an electron pair are vetoed by rejecting electron positron pairs with an invariant mass below 0.14~GeV. We also reject events with more than one lepton candidate.


We exclude events where the missing momentum is not consistent with the presence of a single neutrino from the semileptonic $B$~decay. In particular we impose the event-level selections on the following three properties: missing mass (magnitude of the missing four-momentum) is required to be $M_{miss}^2 <$ 3~GeV$^2$, the polar angle of the missing three-momentum has to lie within $\theta_{miss} \in [0.3, 2.6]$~rad and the absolute value of the total event charge is restricted to $|\sum_i q_i|< 3$.

The MC samples are scaled to the data luminosity and split up into the following components: $B\to X_c\ell\nu_\ell$~signal, $B\to X_u\ell\nu_\ell$ background, the events where the lepton candidate is misidentified (referred to as {\it fakes} or {\it fake leptons}), $b\to c/\bar c\to\ell$ (secondary leptons), and other $B\bar B$~background (the lepton candidate does not belong in any of these categories). The lepton identification at Belle~II is described in~\cite{Milesi:1866}.

\subsection{Signal extraction}

We extract the amount of $B\to X_c\ell\nu_\ell$ signal and background by performing a fit to the binned c.m.\ lepton momentum distribution, separately in the electron and in the muon samples. We use a maximum likelihood technique using Poisson statistics of both real and MC simulated data~\cite{Barlow:1993dm}. The following components are freely floated in this fit: the $b\to c$ signal, $B\bar B$~backgrounds (including $b\to u$, fake and secondary leptons and other $B\bar B$~backgrounds) and the continuum background. The shape in $p^*_\ell$ of the signal and $B\bar B$~backgrounds components are obtained from MC simulation, while the shape of the continuum background is modeled by off-resonance collision data equivalent to \samplesizeOffRes, taken at the c.m.\ energy of \SI{10.52}{\giga\electronvolt}. 


It is necessary to combine all $B\bar B$~background contributions into a single fit component because they have similar shapes in $p^*_\ell$ and the fit would otherwise have difficulties to distinguish them. However, we vary the relative amounts of these background components when evaluating the systematic uncertainty related to the background and repeat the fit with altered compositions of the background.

\begin{figure}[h]
  \centering
  \includegraphics[height=7.8cm]{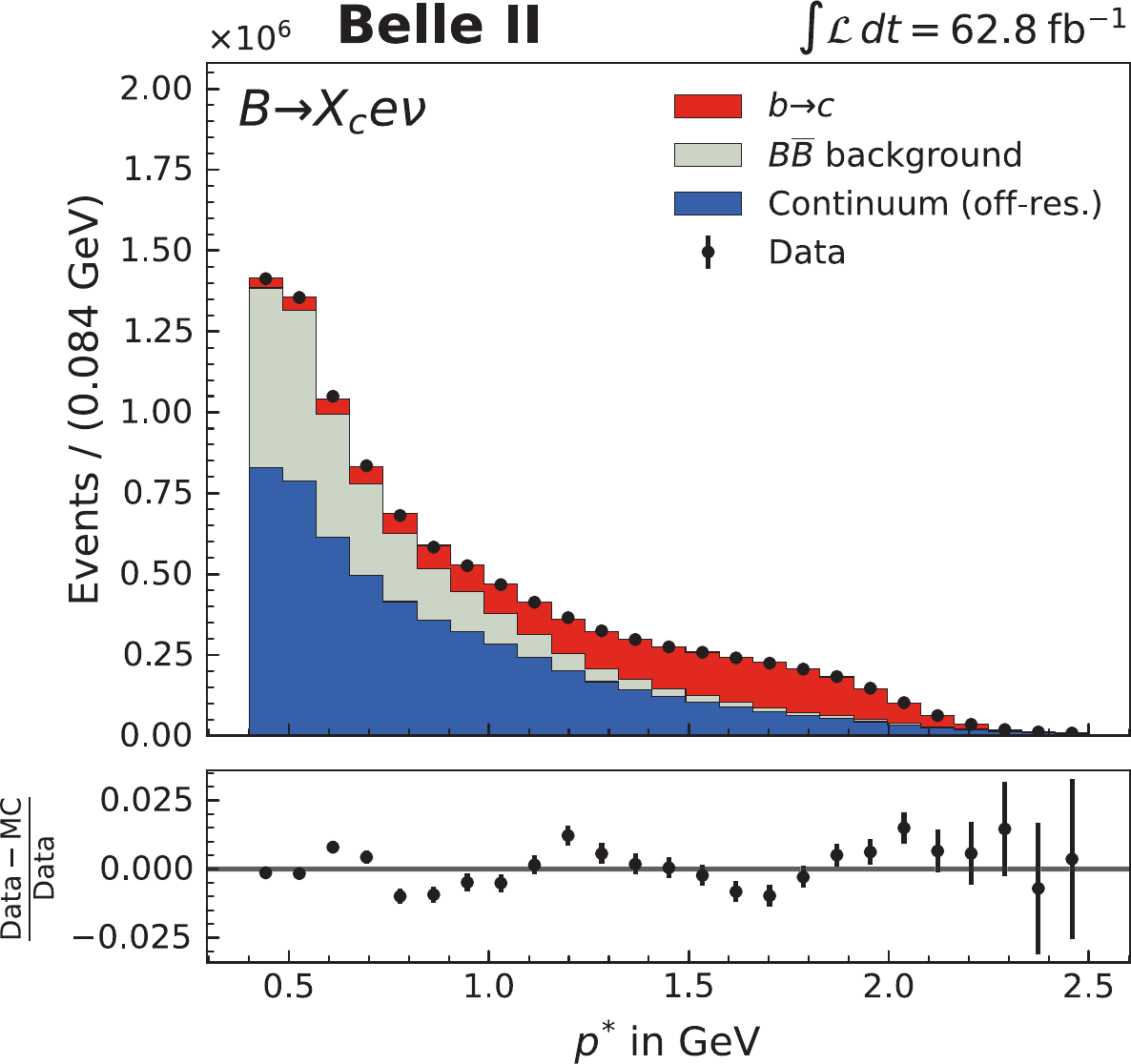}
  \includegraphics[height=7.8cm]{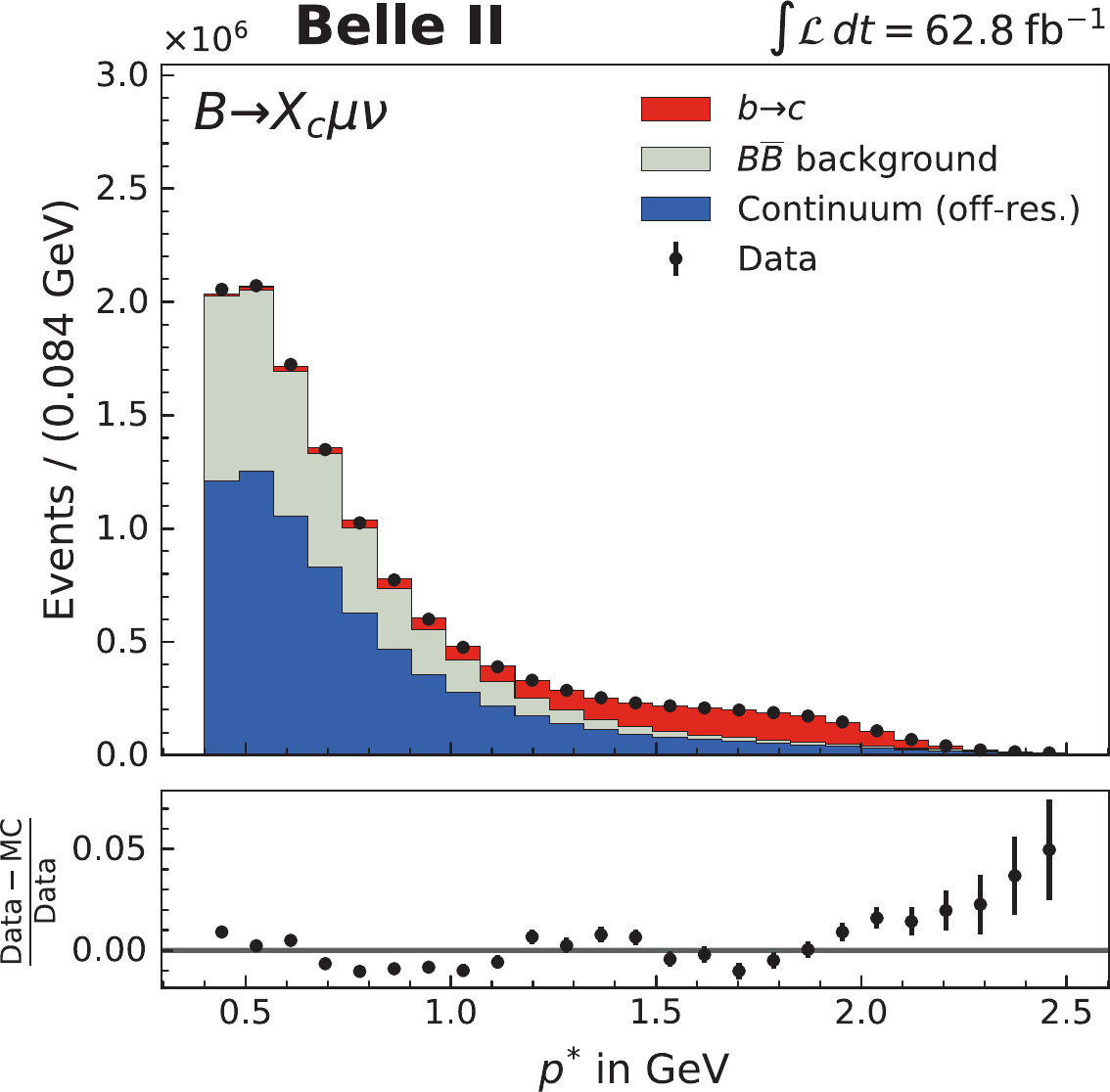}
  \caption{C.m.\ frame electron (left) and muon (right) momentum distributions after the fit. See text for more details.}
  \label{fig:electronMuonPostfit}
\end{figure}

Fig.~\ref{fig:electronMuonPostfit} shows the c.m.\ frame electron and muon momentum distributions after the fit. Table~\ref{table:yieldsFit} gives the yields of the various components and their respective uncertainties determined by the fit.

\clearpage

\begin{table}[ht]
\centering
\caption{Yields in the electron and muon samples. Note that the $b\to u$, fake and secondary leptons and other $B\bar B$~background components are combined in a single fit component and that they are split up here for better understanding. See text for more details.}
\begin{tabular}{| c || c | c |}
    \hline
Yield    & Electron mode & Muon mode \\	\hline \hline	
Signal	&	$(1.932	\pm	0.006)\times 10^6$	&	$(1.501	\pm	0.007)\times 10^6$	\\
$b \to u$ 	background	&	$(53.4	\pm	0.4)\times 10^3$ & $(52	\pm	1)\times 10^3$	\\
Fake leptons		&	$(1.258	\pm	0.009)\times 10^6$	&	$(3.15	\pm	0.07)\times 10^6$	\\
Secondaries		&	$(1.324	\pm	0.009)\times 10^6$	&	$(0.89	\pm	0.02)\times 10^6$	\\
Other $B\bar{B}$ background	&	$(5.42	\pm	0.04)\times 10^3$	&	$(4.33	\pm	0.09)\times 10^3$	\\
Continuum	&	$(5.51	\pm	0.02)\times 10^6$	&	$(7.35	\pm	0.09)\times 10^6$	\\ \hline \hline
Sum 		&	$(10.08	\pm	0.03)\times 10^6$	&	$(13.0	\pm	0.1)\times 10^6$	\\ \hline
    \end{tabular}
\label{table:yieldsFit}
\end{table}


\section{Results and systematic uncertainties} \label{sec:4}

\subsection{Inclusive semileptonic branching fraction}

In this section we determine the inclusive branching fraction of semileptonic decays $B\to X_c\ell\nu_\ell$ where $B$ is a state with the average lifetime of $B^+$ and $B^0$, $\tau=(\tau(B^+)+\tau(B^0))/2=(1.579\pm 0.004)$~ps~\cite{Zyla:2020zbs}. As spectator effects in semileptonic decays are known to be small~\cite{Neubert:1996we,Kirk:2017juj}, we assume a common semileptonic width
\begin{equation}
  \Gamma_\mathrm{s.l.}=\frac{{\cal B}(B^+\to X_c\ell\nu_\ell)}{\tau(B^+)}=\frac{{\cal B}(B^0\to X_c\ell\nu_\ell)}{\tau(B^0)}=\frac{{\cal B}(B\to X_c\ell\nu_\ell)}{\tau}~,
\end{equation}
and calculate the inclusive semileptonic branching fraction as
\begin{equation}
  {\cal B}(B \to X_c \ell \nu_\ell) = \frac{N^\ell_\mathrm{sig}\tau}{2 N_{B\bar{B}}\Big(f_+\epsilon^\ell(B^+)\tau(B^+)+f_0\epsilon^\ell(B^0)\tau(B^0)\Big)},
\end{equation}
where $N^\ell_\mathrm{sig}$ is the fitted number of signal events in the respective sample, $N_{B\bar{B}}$ is the total number of $B\bar{B}$ pairs in the data sample and $\epsilon^\ell(B)$ is the signal selection efficiency in the respective sample. The factor of two accounts for the fact that both $B$~mesons in the $\Upsilon(4S)$~event can contribute to the signal. The factors $\tau(B^+/B^0)$ are the mean lifetimes of the mesons and the $f_{+/0}$ are the production fractions of the two $B$~species at the $\Upsilon(4S)$. We determine them from $f_+/f_0 = 1.058 \pm 0.024$~\cite{Zyla:2020zbs} to be $f_+=0.514\pm 0.006$ and $f_0=0.486\pm 0.006$.

The signal selection efficiencies were determined from MC simulation for the $B^+$ and $B^0$ events separately. The electron mode efficiencies after all applied selections are $\epsilon^e(B^+) = 15.76\% $ and $\epsilon^e(B^0) = 12.40\%$. The muon mode has somewhat lower signal selection efficiencies of $\epsilon^\mu(B^+) = 12.99\% $ and $\epsilon^\mu(B^0) = 10.03\%$.


From this equation we obtain the following branching fractions in the electron and muon samples. The uncertainty is statistical only, {\it i.e.}, corresponds to the uncertainty in the fitted signal fraction.
\begin{eqnarray}
  {\cal B}(B\to X_c e\nu_e) & = & (9.97\pm 0.03(\text{stat}))\%~,\\
  {\cal B}(B\to X_c\mu\nu_\mu) & = & (9.47\pm 0.05(\text{stat}))\%~.
\end{eqnarray}
 
\subsection{Systematic Uncertainties}\label{systematicUnc}


The main contribution is model uncertainty in the $B\to X_c\ell\nu_\ell$ signal and in the $B\bar B$~background component. The \btoxclnu\ modelling uncertainty in Monte Carlo was determined in the following way: At first, the inclusive signal sample was split into 30 separate decay modes. The branching fraction of the mode under consideration was varied by $\pm 1 \sigma$ of the current average branching fraction, taken from the Particle Data Group ~\cite{Zyla:2020zbs}. The whole sample was then fitted again and the number of signal events was obtained from the fit. The systematic uncertainty was calculated for each decay mode from the difference between maximal and minimal yield, and the true signal yield from Table~\ref{table:yieldsFit}. The full modeling uncertainty is calculated by adding the separate contributions in quadrature.


The decay form factors affect the shape of the Monte Carlo template in center-of-mass (c.m.) momentum $p_\ell^*$. The form factor uncertainty is estimated by assuming the Caprini, Lellouch and Neubert (CLN) parameterization~\cite{Caprini:1997mu} for the $B\to D^*\ell\nu_\ell$ and $B\to D\ell\nu_\ell$~decays and varying the form factor parameters within their ranges of uncertainty~\cite{Amhis:2019ckw}.

To estimate the uncertainty in the $B\bar{B}$~background, we vary all four contributions ($b\to u$, secondary leptons, fake leptons and others) by 5\%, which roughly corresponds to the difference between pre-fit and post-fit yields of the background. We determine uncertainties in the same way as for the \btoxclnu\ model. Furthermore, we constrain continuum background to the ratio between on- and off-resonance data of \samplesize/\samplesizeOffRes, allowing it to float only within the uncertainty of the luminosity measurement (later referred to as `fixed' continuum ratio). The uncertainty assigned to the continuum background is the difference between yields with `fixed' and fully floating fraction in the fit. Details on the determination of the background model uncertainty are collected in Table~\ref{table:systematicsBkgs}.
\begin{table}[h]
\centering
\caption{Determination of the background model uncertainty. The table shows the change in fitted signal yield when varying individual background components by $\pm 5\%$.}
\begin{tabular}{| c || c | c | c || c | c | c |}
    \hline
     & \multicolumn{3}{c||}{Electron mode} & \multicolumn{3}{c|}{Muon mode} \\ 
Varying background	&	$N_{sig, 0.95}$	&	$N_{sig, 1.05}$	&	$\sigma_{rel}$ [\%]	&	$N_{sig, 0.95}$	& $N_{sig, 1.05}$ &	$\sigma_{rel}$ [\%]	\\ \hline \hline
$b \to u$	&	1934639	&	1928820	&	0.15	&	1505782	&	1509395	&	0.12	\\
Fake leptons	&	1930435	&	1928590	&	0.05	&	1519457	&	1511576	&	0.26	\\
Secondaries	&	1927430	&	1934593	&	0.19	&	1506999	&	1508290	&	0.04	\\
Others	&	1932781	&	1932126	&	0.02	&	1508798	&	1503859	&	0.16	\\
Continuum data (`fixed' ratio)	& \multicolumn{2}{c|}{1925908} & 0.34 & \multicolumn{2}{c|}{1457793} & 2.91\\ 
    \hline
    \end{tabular}
\label{table:systematicsBkgs}
\end{table}

Other components are uncertainties related to tracking, to the counting of $B\bar B$~events and to lepton identification. The uncertainty related to lepton identification is estimated by generating 200 variations of the simulated events with lepton identification efficiency and misidentification rates chosen randomly within their respective uncertainties. For each variation, the number of signal events ($N_{sig,i}$) is calculated. The mean value and the standard deviation of the distribution of the obtained yields $N_{sig,i}$ are used to determine the lepton identification uncertainty.

A tracking uncertainty of 0.69\% is applied to the only charged particle that is reconstructed. The uncertainty from limited MC sample size in the reconstruction efficiency $\epsilon^\ell(B)$ is at the sub-permille level and therefore negligible. Table~\ref{table:systematics} summarizes our estimate of the systematic uncertainty in the electron and muon samples. The different components of systematic uncertainty are added in quadrature and the overall relative systematic uncertainties are found to be 3.77\% and 4.79\% for the electron and muon modes, respectively.

\begin{table}[h]
\centering
\caption{Estimated relative systematic uncertainty on the \btoxclnu\ branching fraction measurement in the two modes.}
\label{table:systematics}
\begin{tabular}{|c||c|c|}
    \hline
    &  \multicolumn{2}{c|}{Relative uncertainty [\%]} \\
    Contribution & Electron mode  & Muon mode \\ \hline  \hline
Tracking	&	0.69	&	0.69	\\
$N_{B\bar{B}}$	&	1.1	&	1.1	\\
Lepton ID corrections &	1.64	&	2.33	\\
$f_0/f_+$, $B$ lifetime & 1.2 & 1.2 \\
\btoxclnu\ branching fractions	&	2.65 & 2.15	\\
\btoxclnu\ form factors & 1.11 & 1.11 \\
$B\bar B$ background model	&	0.24	&	0.34	\\
Off-resonance data model	&	0.34	&	2.91	\\
    \hline
    \hline
    Sum & 3.77 & 4.79 \\ \hline
\end{tabular}
\\
\end{table}

\section{Conclusion}

We have measured the inclusive \btoxclnu\ branching ratio in a Belle II sample corresponding to \samplesize\ of integrated luminosity. The preliminary results for both lepton modes are
\begin{eqnarray}
  {\cal B}(B\to X_c e\nu_e) & = & (9.97\pm0.03(\text{stat})\pm 0.38(\text{sys}))\%~,\\
  {\cal B}(B\to X_c\mu\nu_\mu) & = & (9.47\pm 0.05(\text{stat})\pm 0.45(\text{sys}))\%~.
\end{eqnarray}

The combined branching fraction is determined as the weighted mean. We conservatively assume electron and muon systematic uncertainties to be fully correlated and use the (larger) muon systematic uncertainty for the combined result. The average semileptonic branching fraction $B\to X_c\ell\nu_\ell$ (where $\ell$ can be either an electron or a muon) is thus found to be
\begin{equation}
  {\cal B}(B\to X_c\ell\nu_\ell) = (9.75\pm 0.03(\text{stat})\pm 0.47(\text{sys}))\%.
\end{equation}


\section{Acknowledgments}
\input{acknowledgements}

\bibliography{references.bib}

\end{document}

%% file: authors-conf2021.tex
\newcommand{\instCPPM}{Aix Marseille Universit\'{e}, CNRS/IN2P3, CPPM, 13288 Marseille, France}
\newcommand{\instYerevan}{Alikhanyan National Science Laboratory, Yerevan 0036, Armenia}
\newcommand{\instBeihang}{Beihang University, Beijing 100191, China}
\newcommand{\instBNL}{Brookhaven National Laboratory, Upton, New York 11973, U.S.A.}
\newcommand{\instBINP}{Budker Institute of Nuclear Physics SB RAS, Novosibirsk 630090, Russian Federation}
\newcommand{\instCMU}{Carnegie Mellon University, Pittsburgh, Pennsylvania 15213, U.S.A.}
\newcommand{\instCinvestavIPN}{Centro de Investigacion y de Estudios Avanzados del Instituto Politecnico Nacional, Mexico City 07360, Mexico}
\newcommand{\instPrague}{Faculty of Mathematics and Physics, Charles University, 121 16 Prague, Czech Republic}
\newcommand{\instChiangMai}{Chiang Mai University, Chiang Mai 50202, Thailand}
\newcommand{\instChiba}{Chiba University, Chiba 263-8522, Japan}
\newcommand{\instChonnam}{Chonnam National University, Gwangju 61186, South Korea}
\newcommand{\instCAU}{Chung-Ang University, Seoul 06974, South Korea}
\newcommand{\instConacyt}{Consejo Nacional de Ciencia y Tecnolog\'{\i}a, Mexico City 03940, Mexico}
\newcommand{\instDESY}{Deutsches Elektronen--Synchrotron, 22607 Hamburg, Germany}
\newcommand{\instDuke}{Duke University, Durham, North Carolina 27708, U.S.A.}
\newcommand{\instITAR}{Institute of Theoretical and Applied Research (ITAR), Duy Tan University, Hanoi 100000, Vietnam}
\newcommand{\instRomaENEA}{ENEA Casaccia, I-00123 Roma, Italy}
\newcommand{\instFuJen}{Department of Physics, Fu Jen Catholic University, Taipei 24205, Taiwan}
\newcommand{\instFudan}{Key Laboratory of Nuclear Physics and Ion-beam Application (MOE) and Institute of Modern Physics, Fudan University, Shanghai 200443, China}
\newcommand{\instGoettingen}{II. Physikalisches Institut, Georg-August-Universit\"{a}t G\"{o}ttingen, 37073 G\"{o}ttingen, Germany}
\newcommand{\instGifu}{Gifu University, Gifu 501-1193, Japan}
\newcommand{\instSOKENDAI}{The Graduate University for Advanced Studies (SOKENDAI), Hayama 240-0193, Japan}
\newcommand{\instGyeongsang}{Gyeongsang National University, Jinju 52828, South Korea}
\newcommand{\instHanyang}{Department of Physics and Institute of Natural Sciences, Hanyang University, Seoul 04763, South Korea}
\newcommand{\instKEK}{High Energy Accelerator Research Organization (KEK), Tsukuba 305-0801, Japan}
\newcommand{\instJPARC}{J-PARC Branch, KEK Theory Center, High Energy Accelerator Research Organization (KEK), Tsukuba 305-0801, Japan}
\newcommand{\instHiroshima}{Hiroshima University, Higashi-Hiroshima, Hiroshima 739-8530, Japan}
\newcommand{\instFrascati}{INFN Laboratori Nazionali di Frascati, I-00044 Frascati, Italy}
\newcommand{\instNapoliINFN}{INFN Sezione di Napoli, I-80126 Napoli, Italy}
\newcommand{\instPadovaINFN}{INFN Sezione di Padova, I-35131 Padova, Italy}
\newcommand{\instPerugiaINFN}{INFN Sezione di Perugia, I-06123 Perugia, Italy}
\newcommand{\instPisaINFN}{INFN Sezione di Pisa, I-56127 Pisa, Italy}
\newcommand{\instRomaINFN}{INFN Sezione di Roma, I-00185 Roma, Italy}
\newcommand{\instRomaTreINFN}{INFN Sezione di Roma Tre, I-00146 Roma, Italy}
\newcommand{\instTorinoINFN}{INFN Sezione di Torino, I-10125 Torino, Italy}
\newcommand{\instTriesteINFN}{INFN Sezione di Trieste, I-34127 Trieste, Italy}
\newcommand{\instIISER}{Indian Institute of Science Education and Research Mohali, SAS Nagar, 140306, India}
\newcommand{\instIITBhubaneswar}{Indian Institute of Technology Bhubaneswar, Satya Nagar 751007, India}
\newcommand{\instIITGuwahati}{Indian Institute of Technology Guwahati, Assam 781039, India}
\newcommand{\instIITHyderabad}{Indian Institute of Technology Hyderabad, Telangana 502285, India}
\newcommand{\instIITMadras}{Indian Institute of Technology Madras, Chennai 600036, India}
\newcommand{\instIndiana}{Indiana University, Bloomington, Indiana 47408, U.S.A.}
\newcommand{\instIHEPRussia}{Institute for High Energy Physics, Protvino 142281, Russian Federation}
\newcommand{\instHEPHYVienna}{Institute of High Energy Physics, Vienna 1050, Austria}
\newcommand{\instIHEPChina}{Institute of High Energy Physics, Chinese Academy of Sciences, Beijing 100049, China}
\newcommand{\instIPP}{Institute of Particle Physics (Canada), Victoria, British Columbia V8W 2Y2, Canada}
\newcommand{\instIOP}{Institute of Physics, Vietnam Academy of Science and Technology (VAST), Hanoi, Vietnam}
\newcommand{\instIFIC}{Instituto de Fisica Corpuscular, Paterna 46980, Spain}
\newcommand{\instISU}{Iowa State University, Ames, Iowa 50011, U.S.A.}
\newcommand{\instJAEA}{Advanced Science Research Center, Japan Atomic Energy Agency, Naka 319-1195, Japan}
\newcommand{\instMainz}{Institut f\"{u}r Kernphysik, Johannes Gutenberg-Universit\"{a}t Mainz, D-55099 Mainz, Germany}
\newcommand{\instGiessen}{Justus-Liebig-Universit\"{a}t Gie\ss{}en, 35392 Gie\ss{}en, Germany}
\newcommand{\instKarlsruhe}{Institut f\"{u}r Experimentelle Teilchenphysik, Karlsruher Institut f\"{u}r Technologie, 76131 Karlsruhe, Germany}
\newcommand{\instKitasato}{Kitasato University, Sagamihara 252-0373, Japan}
\newcommand{\instKISTI}{Korea Institute of Science and Technology Information, Daejeon 34141, South Korea}
\newcommand{\instKoreaUnivKU}{Korea University, Seoul 02841, South Korea}
\newcommand{\instKSU}{Kyoto Sangyo University, Kyoto 603-8555, Japan}
\newcommand{\instKyotoU}{Kyoto University, Kyoto 606-8501, Japan}
\newcommand{\instKyungpook}{Kyungpook National University, Daegu 41566, South Korea}
\newcommand{\instLPI}{P.N. Lebedev Physical Institute of the Russian Academy of Sciences, Moscow 119991, Russian Federation}
\newcommand{\instLNNU}{Liaoning Normal University, Dalian 116029, China}
\newcommand{\instLMU}{Ludwig Maximilians University, 80539 Munich, Germany}
\newcommand{\instLuther}{Luther College, Decorah, Iowa 52101, U.S.A.}
\newcommand{\instMNITJaipur}{Malaviya National Institute of Technology Jaipur, Jaipur 302017, India}
\newcommand{\instMPP}{Max-Planck-Institut f\"{u}r Physik, 80805 M\"{u}nchen, Germany}
\newcommand{\instMPGHLL}{Semiconductor Laboratory of the Max Planck Society, 81739 M\"{u}nchen, Germany}
\newcommand{\instMcGill}{McGill University, Montr\'{e}al, Qu\'{e}bec, H3A 2T8, Canada}
\newcommand{\instMEPhI}{Moscow Physical Engineering Institute, Moscow 115409, Russian Federation}
\newcommand{\instNagoya}{Graduate School of Science, Nagoya University, Nagoya 464-8602, Japan}
\newcommand{\instNagoyaIAR}{Institute for Advanced Research, Nagoya University, Nagoya 464-8602, Japan}
\newcommand{\instNagoyaKMI}{Kobayashi-Maskawa Institute, Nagoya University, Nagoya 464-8602, Japan}
\newcommand{\instNaraWu}{Nara Women's University, Nara 630-8506, Japan}
\newcommand{\instHSE}{National Research University Higher School of Economics, Moscow 101000, Russian Federation}
\newcommand{\instNTUTaiwan}{Department of Physics, National Taiwan University, Taipei 10617, Taiwan}
\newcommand{\instNUUTaiwan}{National United University, Miao Li 36003, Taiwan}
\newcommand{\instKrakow}{H. Niewodniczanski Institute of Nuclear Physics, Krakow 31-342, Poland}
\newcommand{\instNiigata}{Niigata University, Niigata 950-2181, Japan}
\newcommand{\instNSU}{Novosibirsk State University, Novosibirsk 630090, Russian Federation}
\newcommand{\instOkinawa}{Okinawa Institute of Science and Technology, Okinawa 904-0495, Japan}
\newcommand{\instOsakaCity}{Osaka City University, Osaka 558-8585, Japan}
\newcommand{\instRCNP}{Research Center for Nuclear Physics, Osaka University, Osaka 567-0047, Japan}
\newcommand{\instPNNL}{Pacific Northwest National Laboratory, Richland, Washington 99352, U.S.A.}
\newcommand{\instPanjab}{Panjab University, Chandigarh 160014, India}
\newcommand{\instPanjabPAU}{Punjab Agricultural University, Ludhiana 141004, India}
\newcommand{\instRIKENMSL}{Meson Science Laboratory, Cluster for Pioneering Research, RIKEN, Saitama 351-0198, Japan}
\newcommand{\instXavier}{St. Francis Xavier University, Antigonish, Nova Scotia, B2G 2W5, Canada}
\newcommand{\instSeoul}{Seoul National University, Seoul 08826, South Korea}
\newcommand{\instSPU}{Showa Pharmaceutical University, Tokyo 194-8543, Japan}
\newcommand{\instSoochow}{Soochow University, Suzhou 215006, China}
\newcommand{\instSoongsil}{Soongsil University, Seoul 06978, South Korea}
\newcommand{\instLjubljanaJSI}{J. Stefan Institute, 1000 Ljubljana, Slovenia}
\newcommand{\instKyiv}{Taras Shevchenko National Univ. of Kiev, Kiev, Ukraine}
\newcommand{\instTata}{Tata Institute of Fundamental Research, Mumbai 400005, India}
\newcommand{\instTUM}{Department of Physics, Technische Universit\"{a}t M\"{u}nchen, 85748 Garching, Germany}
\newcommand{\instTelAviv}{Tel Aviv University, School of Physics and Astronomy, Tel Aviv, 69978, Israel}
\newcommand{\instToho}{Toho University, Funabashi 274-8510, Japan}
\newcommand{\instTohoku}{Department of Physics, Tohoku University, Sendai 980-8578, Japan}
\newcommand{\instTitech}{Tokyo Institute of Technology, Tokyo 152-8550, Japan}
\newcommand{\instTokyoMetropolitan}{Tokyo Metropolitan University, Tokyo 192-0397, Japan}
\newcommand{\instUAS}{Universidad Autonoma de Sinaloa, Sinaloa 80000, Mexico}
\newcommand{\instNapoliUNIV}{Dipartimento di Scienze Fisiche, Universit\`{a} di Napoli Federico II, I-80126 Napoli, Italy}
\newcommand{\instPadovaUNIV}{Dipartimento di Fisica e Astronomia, Universit\`{a} di Padova, I-35131 Padova, Italy}
\newcommand{\instPerugiaUNIV}{Dipartimento di Fisica, Universit\`{a} di Perugia, I-06123 Perugia, Italy}
\newcommand{\instPisaUNIV}{Dipartimento di Fisica, Universit\`{a} di Pisa, I-56127 Pisa, Italy}
\newcommand{\instRomaTreUNIV}{Dipartimento di Matematica e Fisica, Universit\`{a} di Roma Tre, I-00146 Roma, Italy}
\newcommand{\instTorinoUNIV}{Dipartimento di Fisica, Universit\`{a} di Torino, I-10125 Torino, Italy}
\newcommand{\instTriesteUNIV}{Dipartimento di Fisica, Universit\`{a} di Trieste, I-34127 Trieste, Italy}
\newcommand{\instMontreal}{Universit\'{e} de Montr\'{e}al, Physique des Particules, Montr\'{e}al, Qu\'{e}bec, H3C 3J7, Canada}
\newcommand{\instIJCLab}{Universit\'{e} Paris-Saclay, CNRS/IN2P3, IJCLab, 91405 Orsay, France}
\newcommand{\instIPHC}{Universit\'{e} de Strasbourg, CNRS, IPHC, UMR 7178, 67037 Strasbourg, France}
\newcommand{\instAdelaide}{Department of Physics, University of Adelaide, Adelaide, South Australia 5005, Australia}
\newcommand{\instBonn}{University of Bonn, 53115 Bonn, Germany}
\newcommand{\instUBC}{University of British Columbia, Vancouver, British Columbia, V6T 1Z1, Canada}
\newcommand{\instCincinnati}{University of Cincinnati, Cincinnati, Ohio 45221, U.S.A.}
\newcommand{\instFlorida}{University of Florida, Gainesville, Florida 32611, U.S.A.}
\newcommand{\instHawaii}{University of Hawaii, Honolulu, Hawaii 96822, U.S.A.}
\newcommand{\instHeidelberg}{University of Heidelberg, 68131 Mannheim, Germany}
\newcommand{\instLjubljanaUniLJ}{Faculty of Mathematics and Physics, University of Ljubljana, 1000 Ljubljana, Slovenia}
\newcommand{\instLouisville}{University of Louisville, Louisville, Kentucky 40292, U.S.A.}
\newcommand{\instMalaya}{National Centre for Particle Physics, University Malaya, 50603 Kuala Lumpur, Malaysia}
\newcommand{\instLjubljanaUM}{Faculty of Chemistry and Chemical Engineering, University of Maribor, 2000 Maribor, Slovenia}
\newcommand{\instMelbourne}{School of Physics, University of Melbourne, Victoria 3010, Australia}
\newcommand{\instMississippi}{University of Mississippi, University, Mississippi 38677, U.S.A.}
\newcommand{\instUOM}{University of Miyazaki, Miyazaki 889-2192, Japan}
\newcommand{\instPittsburgh}{University of Pittsburgh, Pittsburgh, Pennsylvania 15260, U.S.A.}
\newcommand{\instUSTC}{University of Science and Technology of China, Hefei 230026, China}
\newcommand{\instSAlabama}{University of South Alabama, Mobile, Alabama 36688, U.S.A.}
\newcommand{\instSCarolina}{University of South Carolina, Columbia, South Carolina 29208, U.S.A.}
\newcommand{\instSydney}{School of Physics, University of Sydney, New South Wales 2006, Australia}
\newcommand{\instUTokyo}{Department of Physics, University of Tokyo, Tokyo 113-0033, Japan}
\newcommand{\instEri}{Earthquake Research Institute, University of Tokyo, Tokyo 113-0032, Japan}
\newcommand{\instIPMU}{Kavli Institute for the Physics and Mathematics of the Universe (WPI), University of Tokyo, Kashiwa 277-8583, Japan}
\newcommand{\instVictoria}{University of Victoria, Victoria, British Columbia, V8W 3P6, Canada}
\newcommand{\instVPI}{Virginia Polytechnic Institute and State University, Blacksburg, Virginia 24061, U.S.A.}
\newcommand{\instWayneState}{Wayne State University, Detroit, Michigan 48202, U.S.A.}
\newcommand{\instYamagata}{Yamagata University, Yamagata 990-8560, Japan}
\newcommand{\instYonsei}{Yonsei University, Seoul 03722, South Korea}
\affiliation{\instCPPM}
\affiliation{\instYerevan}
\affiliation{\instBeihang}
\affiliation{\instBNL}
\affiliation{\instBINP}
\affiliation{\instCMU}
\affiliation{\instCinvestavIPN}
\affiliation{\instPrague}
\affiliation{\instChiangMai}
\affiliation{\instChiba}
\affiliation{\instChonnam}
\affiliation{\instConacyt}
\affiliation{\instDESY}
\affiliation{\instDuke}
\affiliation{\instITAR}
\affiliation{\instRomaENEA}
\affiliation{\instFuJen}
\affiliation{\instFudan}
\affiliation{\instGoettingen}
\affiliation{\instGifu}
\affiliation{\instSOKENDAI}
\affiliation{\instGyeongsang}
\affiliation{\instHanyang}
\affiliation{\instKEK}
\affiliation{\instJPARC}
\affiliation{\instHiroshima}
\affiliation{\instFrascati}
\affiliation{\instNapoliINFN}
\affiliation{\instPadovaINFN}
\affiliation{\instPerugiaINFN}
\affiliation{\instPisaINFN}
\affiliation{\instRomaINFN}
\affiliation{\instRomaTreINFN}
\affiliation{\instTorinoINFN}
\affiliation{\instTriesteINFN}
\affiliation{\instIISER}
\affiliation{\instIITBhubaneswar}
\affiliation{\instIITGuwahati}
\affiliation{\instIITHyderabad}
\affiliation{\instIITMadras}
\affiliation{\instIndiana}
\affiliation{\instIHEPRussia}
\affiliation{\instHEPHYVienna}
\affiliation{\instIHEPChina}
\affiliation{\instIPP}
\affiliation{\instIOP}
\affiliation{\instIFIC}
\affiliation{\instISU}
\affiliation{\instJAEA}
\affiliation{\instMainz}
\affiliation{\instGiessen}
\affiliation{\instKarlsruhe}
\affiliation{\instKitasato}
\affiliation{\instKISTI}
\affiliation{\instKoreaUnivKU}
\affiliation{\instKSU}
\affiliation{\instKyotoU}
\affiliation{\instKyungpook}
\affiliation{\instLPI}
\affiliation{\instLNNU}
\affiliation{\instLMU}
\affiliation{\instLuther}
\affiliation{\instMNITJaipur}
\affiliation{\instMPP}
\affiliation{\instMPGHLL}
\affiliation{\instMcGill}
\affiliation{\instMEPhI}
\affiliation{\instNagoya}
\affiliation{\instNagoyaIAR}
\affiliation{\instNagoyaKMI}
\affiliation{\instNaraWu}
\affiliation{\instHSE}
\affiliation{\instNTUTaiwan}
\affiliation{\instNUUTaiwan}
\affiliation{\instKrakow}
\affiliation{\instNiigata}
\affiliation{\instNSU}
\affiliation{\instOkinawa}
\affiliation{\instOsakaCity}
\affiliation{\instRCNP}
\affiliation{\instPNNL}
\affiliation{\instPanjab}
\affiliation{\instPanjabPAU}
\affiliation{\instRIKENMSL}
\affiliation{\instXavier}
\affiliation{\instSeoul}
\affiliation{\instSPU}
\affiliation{\instSoochow}
\affiliation{\instSoongsil}
\affiliation{\instLjubljanaJSI}
\affiliation{\instKyiv}
\affiliation{\instTata}
\affiliation{\instTUM}
\affiliation{\instTelAviv}
\affiliation{\instToho}
\affiliation{\instTohoku}
\affiliation{\instTitech}
\affiliation{\instTokyoMetropolitan}
\affiliation{\instUAS}
\affiliation{\instNapoliUNIV}
\affiliation{\instPadovaUNIV}
\affiliation{\instPerugiaUNIV}
\affiliation{\instPisaUNIV}
\affiliation{\instRomaTreUNIV}
\affiliation{\instTorinoUNIV}
\affiliation{\instTriesteUNIV}
\affiliation{\instMontreal}
\affiliation{\instIJCLab}
\affiliation{\instIPHC}
\affiliation{\instAdelaide}
\affiliation{\instBonn}
\affiliation{\instUBC}
\affiliation{\instCincinnati}
\affiliation{\instFlorida}
\affiliation{\instHawaii}
\affiliation{\instHeidelberg}
\affiliation{\instLjubljanaUniLJ}
\affiliation{\instLouisville}
\affiliation{\instMalaya}
\affiliation{\instLjubljanaUM}
\affiliation{\instMelbourne}
\affiliation{\instMississippi}
\affiliation{\instUOM}
\affiliation{\instPittsburgh}
\affiliation{\instUSTC}
\affiliation{\instSAlabama}
\affiliation{\instSCarolina}
\affiliation{\instSydney}
\affiliation{\instUTokyo}
\affiliation{\instEri}
\affiliation{\instIPMU}
\affiliation{\instVictoria}
\affiliation{\instVPI}
\affiliation{\instWayneState}
\affiliation{\instYamagata}
\affiliation{\instYonsei}
  \author{F.~Abudin{\'e}n}\affiliation{\instTriesteINFN} 
  \author{I.~Adachi}\affiliation{\instKEK}\affiliation{\instSOKENDAI} 
  \author{R.~Adak}\affiliation{\instFudan} 
  \author{K.~Adamczyk}\affiliation{\instKrakow} 
  \author{L.~Aggarwal}\affiliation{\instPanjab} 
  \author{P.~Ahlburg}\affiliation{\instBonn} 
  \author{H.~Ahmed}\affiliation{\instXavier} 
  \author{J.~K.~Ahn}\affiliation{\instKoreaUnivKU} 
  \author{H.~Aihara}\affiliation{\instUTokyo} 
  \author{N.~Akopov}\affiliation{\instYerevan} 
% ...
\collaboration{Belle II Collaboration}

%% file: acknowledgements.tex
We thank the SuperKEKB group for the excellent operation of the
accelerator; the KEK cryogenics group for the efficient
operation of the solenoid; the KEK computer group for
on-site computing support; and the raw-data centers at
BNL, DESY, GridKa, IN2P3, and INFN for off-site computing support.
This work was supported by the following funding sources:
Science Committee of the Republic of Armenia Grant No. 20TTCG-1C010;
Australian Research Council and research grant Nos.
DP180102629, 
DP170102389, 
DP170102204, 
DP150103061, 
FT130100303, 
FT130100018,
and
FT120100745;
Austrian Federal Ministry of Education, Science and Research,
Austrian Science Fund No. P 31361-N36, and
Horizon 2020 ERC Starting Grant no. 947006 ``InterLeptons''; 
Natural Sciences and Engineering Research Council of Canada, Compute Canada and CANARIE;
Chinese Academy of Sciences and research grant No. QYZDJ-SSW-SLH011,
National Natural Science Foundation of China and research grant Nos.
11521505,
11575017,
11675166,
11761141009,
11705209,
and
11975076,
LiaoNing Revitalization Talents Program under contract No. XLYC1807135,
Shanghai Municipal Science and Technology Committee under contract No. 19ZR1403000,
Shanghai Pujiang Program under Grant No. 18PJ1401000,
and the CAS Center for Excellence in Particle Physics (CCEPP);
the Ministry of Education, Youth and Sports of the Czech Republic under Contract No.~LTT17020 and 
Charles University grants SVV 260448 and GAUK 404316;
European Research Council, 7th Framework PIEF-GA-2013-622527, 
Horizon 2020 ERC-Advanced Grants No. 267104 and 884719,
Horizon 2020 ERC-Consolidator Grant No. 819127,
Horizon 2020 Marie Sklodowska-Curie grant agreement No. 700525 `NIOBE,' 
and
Horizon 2020 Marie Sklodowska-Curie RISE project JENNIFER2 grant agreement No. 822070 (European grants);
L'Institut National de Physique Nucl\'{e}aire et de Physique des Particules (IN2P3) du CNRS (France);
BMBF, DFG, HGF, MPG, and AvH Foundation (Germany);
Department of Atomic Energy under Project Identification No. RTI 4002 and Department of Science and Technology (India);
Israel Science Foundation grant No. 2476/17,
United States-Israel Binational Science Foundation grant No. 2016113, and
Israel Ministry of Science grant No. 3-16543;
Istituto Nazionale di Fisica Nucleare and the research grants BELLE2;
Japan Society for the Promotion of Science,  Grant-in-Aid for Scientific Research grant Nos.
16H03968, 
16H03993, 
16H06492,
16K05323, 
17H01133, 
17H05405, 
18K03621, 
18H03710, 
18H05226,
19H00682, 
26220706,
and
26400255,
the National Institute of Informatics, and Science Information NETwork 5 (SINET5), 
and
the Ministry of Education, Culture, Sports, Science, and Technology (MEXT) of Japan;  
National Research Foundation (NRF) of Korea Grant Nos.
2016R1\-D1A1B\-01010135,
2016R1\-D1A1B\-02012900,
2018R1\-A2B\-3003643,
2018R1\-A6A1A\-06024970,
2018R1\-D1A1B\-07047294,
2019K1\-A3A7A\-09033840,
and
2019R1\-I1A3A\-01058933,
Radiation Science Research Institute,
Foreign Large-size Research Facility Application Supporting project,
the Global Science Experimental Data Hub Center of the Korea Institute of Science and Technology Information
and
KREONET/GLORIAD;
Universiti Malaya RU grant, Akademi Sains Malaysia and Ministry of Education Malaysia;
Frontiers of Science Program contracts
FOINS-296,
CB-221329,
CB-236394,
CB-254409,
and
CB-180023, and SEP-CINVESTAV research grant 237 (Mexico);
the Polish Ministry of Science and Higher Education and the National Science Center;
the Ministry of Science and Higher Education of the Russian Federation,
Agreement 14.W03.31.0026, and
the HSE University Basic Research Program, Moscow;
University of Tabuk research grants
S-0256-1438 and S-0280-1439 (Saudi Arabia);
Slovenian Research Agency and research grant Nos.
J1-9124
and
P1-0135; 
Agencia Estatal de Investigacion, Spain grant Nos.
FPA2014-55613-P
and
FPA2017-84445-P,
and
CIDEGENT/2018/020 of Generalitat Valenciana;
Ministry of Science and Technology and research grant Nos.
MOST106-2112-M-002-005-MY3
and
MOST107-2119-M-002-035-MY3, 
and the Ministry of Education (Taiwan);
Thailand Center of Excellence in Physics;
TUBITAK ULAKBIM (Turkey);
National Research Foundation of Ukraine, project No. 2020.02/0257,
and
Ministry of Education and Science of Ukraine;
the US National Science Foundation and research grant Nos.
PHY-1807007 
and
PHY-1913789, 
and the US Department of Energy and research grant Nos.
DE-AC06-76RLO1830, 
DE-SC0007983, 
DE-SC0009824, 
DE-SC0009973, 
DE-SC0010073, 
DE-SC0010118, 
DE-SC0010504, 
DE-SC0011784, 
DE-SC0012704, 
DE-SC0021274; 
and
the Vietnam Academy of Science and Technology (VAST) under grant DL0000.05/21-23.